\documentstyle[12pt]{article}

\begin{document}

\title{How a Non-hierarchical Neutrino Mass Matrix Can Arise}

\author{S.M. Barr\\Bartol Research Institute\\
University of Delaware\\Newark, DE 19716}

\date{}
\maketitle

\begin{abstract}

One puzzle of neutrino masses and mixings is that they do not exhibit
the kind of strong ``hierarchy" that is found for the quarks and charged
leptons. Neutrino mass ratios and mixing angles are not small. A
possible reason for this is proposed here. It is based on the fact that
typical realistic grand unified models contain particles with 
unification-scale masses which, when integrated out, can yield a
neutrino mass matrix that is not of the standard seesaw form.

\end{abstract}

In the standard (or type I) see-saw mechanism \cite{seesaw}, 
the observed left-handed
neutrinos derive their masses from their Dirac couplings to right-handed
neutrinos. These right-handed neutrinos have Majorana masses that in grand 
unified theories (GUTs) are expected to be of order the unified scale
$M_{GUT}$. When the right-handed neutrinos are integrated out, an effective
$d=5$ operator is induced that produces a mass matrix for the left-handed 
neutrinos given by the well-known ``see-saw formula":
\begin{equation}
M_{\nu} = - M_N M_R^{-1} M_N^T.
\end{equation}

\noindent
Here $M_N$ is the Dirac mass matrix of the neutrinos, whose elements
are less than or of order the Weak scale. $M_R$ is the Majorana
mass matrix of the right-handed neutrinos. And $M_{\nu}$ is the effective
mass matrix of the light left-handed neutrinos. 

In GUTs based on $SO(10)$ or related groups, there is typically a close
connection between the Dirac mass matrix of the neutrinos $M_N$ and 
the (Dirac) mass matrices of the up quarks, down quarks, and charged
leptons ($M_U$, $M_D$, and $M_L$ respectively). For instance, in minimal
$SO(10)$ these matrices are all proportional. The small interfamily 
mass ratios and small CKM angles suggest that these other mass matrices
are ``hierarchical" in structure (i.e. their entries decrease as one
goes up and to the left in the matrix, in the usual convention). 
This would lead us to expect a hierarchical
structure also for $M_N$. In that case, the see-saw 
formula would make $M_{\nu}$ hierarchical unless there were a 
``Dirac-Majorana conspiracy" \cite{bd}, i.e. unless there were hierarchies in 
both $M_N$ and $M_R$ that practically cancelled each other out in 
$M_{\nu}$. Such a conspiracy is not easy to achieve in a natural way. 

On the other hand, the experimental evidence suggests that the light neutrinos
have at most a very weak hierarchy. The solar mixing angle
and atmospheric mixing angle are large, and the ratio of neutrino masses
$m_2/m_3$ is not very small either, about $\frac{1}{6}$ 
judging from the fact that 
$\delta m^2_{atm} \sim 3 \times 10^{-3}$ eV \cite{atm} and 
$\delta m^2_{atm} \sim 7.3 \times 10^{-5}$ eV \cite{solar}. 
Thus we have a problem: if $M_{\nu}$ is given
by the standard see-saw formula, why doesn't it exhibit a strong interfamily 
hierarchy? In this letter we suggest a possible solution. 

In realistic grand unified models, it is typical that there are many 
fermions with masses of order $M_{GUT}$ besides the three right-handed 
neutrinos. When integrated out, these other superheavy fermions
can give effective higher-dimension Yukawa operators that 
contribute to the masses of the observed quarks and leptons. 
In fact, in many models, this is precisely the way in which the quarks 
and leptons of the lighter families get their small masses, i.e. the way
the interfamily mass hierarchies arise. (See \cite{fn,models} for examples).
The crucial point for the present discussion is that these superheavy 
fermion multiplets typically contain neutrino-like states; and these, when
integrated out, give contributions to the masses of the observed light 
neutrinos that are not necessarily of the standard (or ``type I") see-saw form 
shown in Eq. (1). 
These non-type-I contributions to $M_{\nu}$ do not involve the 
effective Dirac neutrino mass matrix $M_N$ and consequently
there is no reason why they have to be 
hierarchical. 

The possibility of non-type-I contributions to $M_{\nu}$ arising from 
integrating out superheavy fermion multiplets in grand unified models was
analyzed in a general way in Ref. \cite{gen}. Here we present
a concrete model that illustrates in a transparent way how a 
non-hierarchical $M_{\nu}$ can arise from these contributions even when family 
symmetry causes $M_N$, $M_U$, $M_D$, and $M_L$ to be hierarchical. 

Consider an $SO(10)$ model in which, in addition to the three families 
(contained in three spinor multiplets, denoted ${\bf 16}_i$, $i=1,2,3$), 
there are three family-antifamily pairs, denoted ${\bf 16}'_i$, 
$\overline{{\bf 16}}'_i$, $i=1,2,3$. There is a vector multiplet of Higgs 
fields, denoted ${\bf 10}_H$, which contains the MSSM doublets of 
Higgs, $H_u$ and $H_d$. There are three Higgs multiplets denoted $\Omega_i$, 
$i=1,2,3$, which contain components that are neutral under the
Standard Model gauge group, have $B = L =0$, and obtain $O(M_{GUT})$ vacuum 
expectation values (VEVs). These will break a $U(1)$ family symmetry 
and play the role of ``familons" or ``Froggatt-Nielson fields" \cite{fn}. 

Under the family symmetry, called $U(1)_F$, the 
multiplets just itemized have the following charges. 
The ${\bf 16}_i$ and $\overline{{\bf 16}}'_i$ have 
charge $q_i$, and the ${\bf 16}'_i$ has charge $-q_i$, 
where $q_1= y$, $q_2 = z$, and $q_3 =0$. The Higgs 
multiplet ${\bf 10}_H$ is neutral under $U(1)_F$, 
and the Froggatt-Nielson fields $\Omega_1$, $\Omega_2$,
and $\Omega_3$ have charges $(-y-z)$, $-z$, and $0$, 
respectively.

These assignments allow the following renormalizable
terms to appear in the Yukawa part of the superpotential.
\begin{equation}
\begin{array}{ccl}
W_{Yukawa} & = & \sum_i M_i (\overline{{\bf 16}}'_i
{\bf 16}'_i ) + ( \overline{{\bf 16}}'_3 {\bf 16}'_3 )
\Omega_3 \\ \\
& + & \sum_i ( {\bf 16}_i {\bf 16}'_i ) {\bf 10}_H
+ ( {\bf 16}_3 {\bf 16}_3 ) {\bf 10}_H
+ ( {\bf 16}'_3 {\bf 16}'_3 ) {\bf 10}_H \\ \\
& + & ( \overline{{\bf 16}}'_3 {\bf 16}_2 +
\overline{{\bf 16}}'_2 {\bf 16}_3 ) \Omega_2 +
 ( \overline{{\bf 16}}'_2 {\bf 16}_1 +
\overline{{\bf 16}}'_1 {\bf 16}_2 ) \Omega_1 + (\overline{{\bf 16}}'_3
{\bf 16}_3 ) \Omega_3. 
\end{array}
\end{equation}

\noindent
We do not show the dimensionless Yukawa couplings, which we assume all 
to be $O(1)$. Note that the explicit mass terms and the couplings of the 
${\bf 10}_H$ are both flavor-diagonal. The non-trivial flavor structure in the 
model comes from the couplings of $\Omega_i$

In addition to all of these multiplets, we require there 
to be Higgs that break $B-L$ at the GUT scale. We take these to be 
in several $\overline{{\bf 126}}_H$ and ${\bf 126}_H$ 
multiplets. (There are reasons, including not having the unified
coupling blow up as it runs from $M_{GUT}$ to $M_{P \ell}$, that it might
be more realistic to assume that $B-L$ is broken by spinors 
of Higgs fields. However, for simplicity of exposition, we assume that 
it is broken by $\overline{{\bf 126}}_H$ and ${\bf 126}_H$. 
It would be quite straightforward to modify the model to replace the 
$\overline{{\bf 126}}_H$ and ${\bf 126}_H$ by spinor Higgs.) 
The mass matrix $M_R$ of the right-handed neutrinos comes 
from terms of the form $( {\bf 16}_i {\bf 16}_j ) 
\overline{{\bf 126}}_H$. We will also assume that a mass matrix 
$\overline{M}'_R$ arises from terms of the form 
$( \overline{{\bf 16}}'_i \overline{{\bf 16}}'_j ) 
{\bf 126}_H$. The $\overline{{\bf 126}}_H$ and 
${\bf 126}_H$ multiplets will have whatever $U(1)_F$
charges are required to allow these couplings.
 
The family hierarchy is assumed to arise from a 
hierarchy in the VEVs of the familons, with
\begin{equation}
\Omega_1 \ll \Omega_2 \ll \Omega_3.
\end{equation}

\noindent
Before considering neutrino masses, let us see how
the hierarchies in $M_U$, $M_D$, and $M_L$ arise.
For specificity, let us consider $M_L$. The multiplets
${\bf 16}_i$, ${\bf 16}'_i$, and $\overline{{\bf 16}}'_i$ contain 
charged leptons that we will denote $(\ell^+_i, \ell^-_i)$, 
$(\ell^{+ \prime}_i, \ell^{- \prime}_i)$, and
$(\overline{\ell^+}'_i, \overline{\ell^-}'_i)$, 
respectively. If the mixing of the ${\bf 16}_i$ and ${\bf 16}'_i$
is small, then it is approximately the case that the
light leptons are contained in the ``unprimed multiplets" ${\bf 16}_i$
while the superheavy leptons are contained in the
``primed multiplets" $\overline{{\bf 16}}'_i$ and
${\bf 16}'_i$. There is a $d=4$ operator that contributes
directly to $M_L$, namely $( \ell^-_3 \ell^+_3) H_d$,
coming from the term $( {\bf 16}_3 {\bf 16}_3) 
{\bf 10}_H$ in Eq. (2). In addition, $M_L$ receives
contributions from integrating out the superheavy
(primed) leptons, as follows.

The terms in Eq. (2) include leptons mass terms of
the form
\begin{equation}
\begin{array}{ccl}
W_{lepton} & \supset & \sum_{ij} M_{ij} 
(\overline{\ell^-}'_i \ell^{- \prime}_j +  
\overline{\ell^+}'_i \ell^{+ \prime}_j ) \\ \\
& + & \sum_{ij} (M_{\Omega})_{ij} 
(\overline{\ell^-}'_i \ell^-_j +  
\overline{\ell^+}'_i \ell^+_j ) \\ \\
& + & \sum_{ij} (m_H)_{ij} (\ell^-_i 
\ell^{+ \prime}_j + \ell^-_i 
\ell^{+ \prime}_j).
\end{array}
\end{equation}

\noindent
where these matrices have the forms
\begin{equation}
M = \left( \begin{array}{ccc}
M_1 & 0 & 0 \\ 0 & M_2 & 0 \\
0 & 0 & M_3 \end{array} \right),\;
M_{\Omega} = \left( \begin{array}{ccc}
0 & \omega_1 & 0 \\ \omega_1 & 0 & \omega_2 \\
0 & \omega_2 & \omega_3 \end{array} \right),
m_H = \left( \begin{array}{ccc}
m_1 & 0 & 0 \\ 0 & m_2 & 0 \\
0 & 0 & m_3 \end{array} \right).
\end{equation}

\noindent
The matrices $M$ and $M_{\Omega}$ have elements of $O(M_{GUT})$, while 
the elements of $m_H$ are of order the weak scale. And $\omega_1 \ll
\omega_2 \ll \omega_3$, $M_1 \sim M_2 \sim M_3$, $m_1 \sim m_2 \sim m_3$. 
When the superheavy, primed leptons are integrated out, the diagrams 
of Fig. 1 lead to a contribution to $M_L$ of the form
\begin{equation}
\begin{array}{ccl}
\delta M_L & = & m_H M^{-1} M_{\Omega} + 
{\rm Transpose} \\ \\
& = & \left( \begin{array}{ccc}
0 & \overline{\omega}_1 & 0 \\
\overline{\omega}_1 & 0 & \overline{\omega}_2 \\
0 & \overline{\omega}_2 & \overline{\omega}_3
\end{array} \right),
\end{array}
\end{equation}

\noindent
where $\overline{\omega}_1 = \omega_1 (\frac{m_1}{M_1} + \frac{m_2}{M_2})$, 
$\overline{\omega}_2 = \omega_2 (\frac{m_2}{M_2} + \frac{m_3}{M_3})$, 
and $\overline{\omega}_3 = \omega_3 (2 \frac{m_3}{M_3})$. Note the 
hierarchical form of $\delta M_L$. The $d=4$ operator 
$\ell^-_3 \ell^+_3 H_d$ just gives a contribution to the 33 element 
that is of the same order as the 33 element of $\delta M_L$.

Since the same types of diagrams give the mass matrices $M_U$, $M_D$, and 
$M_N$, these matrices are all related. In fact, if the 
fields $\Omega_i$ were $SO(10)$ singlets, these matrices would
not ``know" that $SO(10)$ is broken, and one would
have the minimal $SO(10)$ relations $M_L = M_D
\propto M_U = M_N$. If, on the other hand, some of the
$\Omega_i$ are adjoints or other non-singlets, then Clebsch coefficients 
can appear that distinguish the matrices of fermions of different 
types (i.e. of different Standard Model quantum numbers). 

The Dirac mass matrix $M_N$ thus arises both from the direct $d=4$ term
$(\nu_3 N^c_3) H_u$ (from $({\bf 16}_3 {\bf 16}_3) {\bf 10}_H$)
and from the diagrams of Fig. 2. Like $M_L$, it is
hierarchical in form. The right-handed neutrinos $N^c_i$ get 
superlarge Majorana mass matrix $(M_R)_{ij} (N^c_i N^c_j)$
from terms of the form $( {\bf 16}_i {\bf 16}_j ) \overline{{\bf 126}}_H$. 
Consequently, one has the type I see-saw contribution 
$M_{\nu}^{type-I} = - M_N M_R^{-1} M_N^T$, which arises from graphs of 
the kind shown in Fig. 3. However, these are not the only kind of
graphs that can contribute to $M_{\nu}$. If there are terms of the 
form $(\overline{{\bf 16}}'_i \overline{{\bf 16}}'_j) {\bf 126}_H$, 
then a simpler graph, shown in Fig. 4, contributes. 
Basically, this graph ``short circuits" the usual type-I diagrams by 
eliminating the standard right-handed neutrinos $N^c_i$ as intermediate
particles. To put it another way, it ``eliminates the middle-man".
This graph gives 
\begin{equation} 
M_{\nu}^{non-type-I} = m_H M^{-1 T} \overline{M}'_R M^{-1} m_H^T.
\end{equation}

\noindent
Note that the graph in Fig. 4 does {\it not} contain insertions of 
the VEVs of the Froggatt-Nielson fields $\Omega_i$, and therefore 
does not ``know" about the family hierarchy. Consequently, this 
non-type-I contribution to $M_{\nu}$ has no reason to be hierarchical 
and will not in general be so. Moreover, since these non-type-I
contributions do not depend on the small parameters
$\Omega_1/\Omega_3$ and $\Omega_2/\Omega_3$, they will
dominate over the type-I contributions, except perhaps 
for the 33 element where they should be comparable.
One expects, therefore, neutrino mixing angles and neutrino mass 
ratios that are of order one, as is indeed observed. 

It should be noted that if the term $({\bf 16}_i {\bf 16}_j) 
\overline{{\bf 126}}_H$ does not exist, i.e. the matrix $M_R = 0$,
then the form of $M_{\nu}$ is not well approximated by Eq. (7).
The diagrams can be misleading, and one must really do the matrix 
diagonalization to find out the form of $M_{\nu}$. However, if both
$M_R$ and $\overline{M}'_R$ are non-zero and $M_R$ is non-singular,
then $M_{\nu}$ is given aproximately by the sum of Eq. (1) and
Eq.(7).

One might ask whether the even simpler kind of diagram shown in Fig. 5
might give a non-type-I, hierarchical contribution to $M_{\nu}$ if
there were terms of the form $({\bf 16}' {\bf 16}') \overline{{\bf 126}}_H$.
It would appear so from the structure of the diagram; but it turns out
that no non-type-I contribution of this kind arises unless there is
also a contribution of the kind given in Fig. 4 and Eq. (7). This is
not obvious from inspection of the diagrams, but can be shown by
looking in detail at the full neutrino mass matrix diagonalization problem,
as was done in the general analysis of Ref. \cite{gen}.

In conclusion, we see that in realistic GUT models there can be additional
contributions to $M_{\nu}$ that are not of the type I form, and that these
contributions can be non-hierarchical even if a family symmetry enforces
a hierarchical pattern for all the Dirac mass matrices, that of the
neutrinos ($M_N$) and those of the up quarks, down quarks, and charged
leptons ($M_U$, $M_D$, $M_L$). This is due to the fact that there typically
exist in realistic GUT models superheavy neutrino-like fermions 
besides the three
right-handed neutrinos $N^c$ usually taken into account. Integrating
out these other states can induce contributions to $M_{\nu}$ that do not
involve the $N^c$ at all, and thus do not depend on the form of the
Dirac neutrino mass matrix $M_N$. Finally, it should be mentioned that a recent 
paper of Nir and Shadmi \cite{ns} has interesting points of contact with the
present work. For a discussion of the similarities and differences of their
approach and the ideas on which the present work is based, see the discussion
at the end of \cite{gen}.

\newpage

\noindent
{\bf\large Figure Captions}

\vspace{1cm}

\noindent
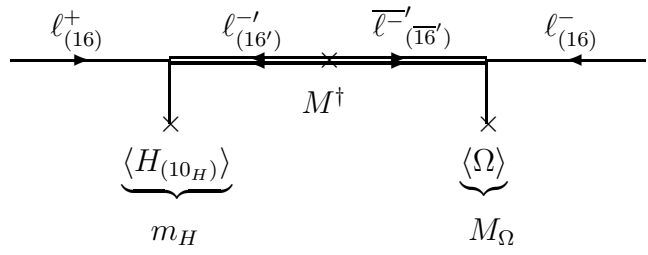
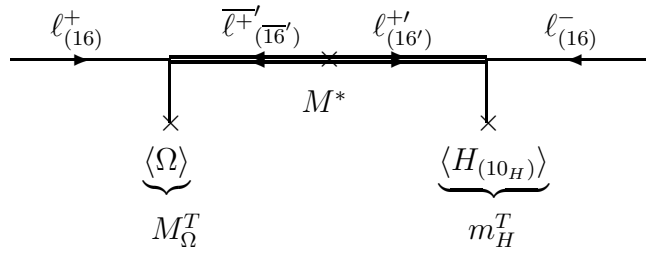
{\bf Fig. 1:}  A contribution to the charged lepton mass matrix $M_L$
coming from integrating out the $O(M_{GUT})$ charged leptons in the ``primed"
multiplets (denoted by the double lines). The coupling to the familons 
$\Omega$ gives the family hierarchy. 

\vspace{0.2cm}

\noindent
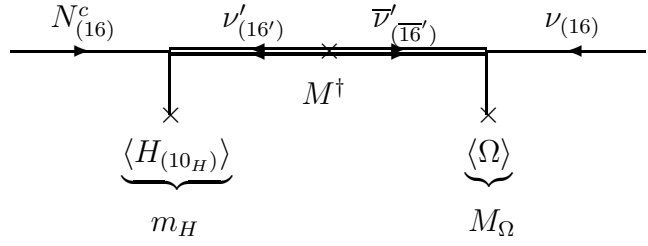
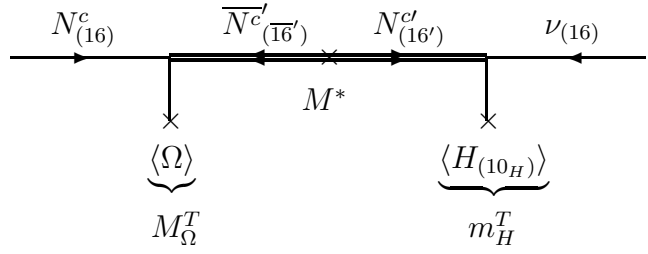
{\bf Fig. 2:}  A contribution to the neutrino Dirac mass $M_N$ coming from
integrating out the $O(M_{GUT})$ neutrinos in the ``primed"
multiplets (denoted by double lines). These diagrams are related by 
$SO(10)$ symmetry to those in Fig. 1.

\vspace{0.2cm}

\noindent
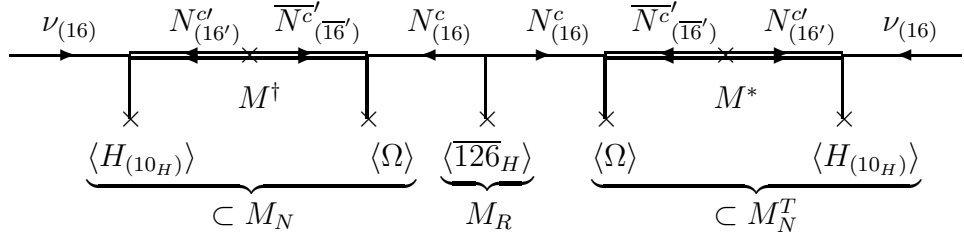
{\bf Fig. 3:} A typical standard (or type I) seesaw contribution to
$M_{\nu}$. The double lines represent the $O(M_{GUT})$ fermions
that are integrated out to generate $M_N$. (See Fig. 2.) 

\vspace{0.2cm}

\noindent
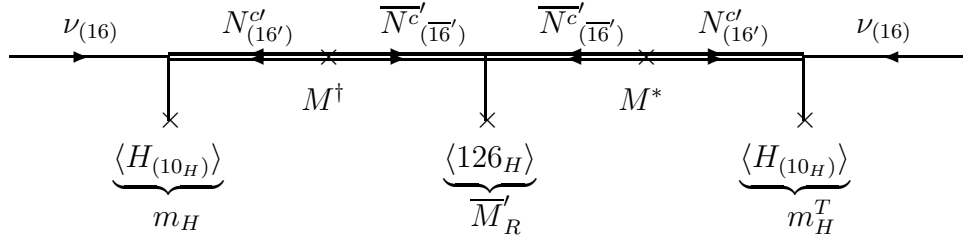
{\bf Fig. 4:} A contribution to $M_{\nu}$ that is not of the
standard type I form. It does not involve the standard right-handed
neutrinos $N^c_i$ of the three families. It also does not involve
the familon field $\Omega$ and is not therefore hierarchical in
structure.

\vspace{0.2cm}

\noindent
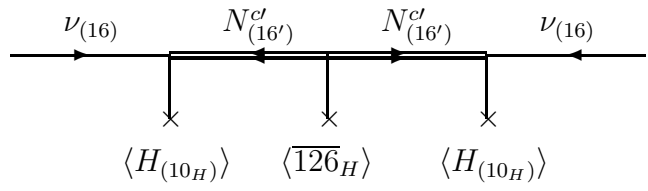
{\bf Fig. 5:} A diagram that seems to give a non-type-I contribution
to $M_{\nu}$ but generally does not.

\newpage 

\begin{picture}(360,216)
\thicklines
\put(60,108){\vector(1,0){30}}
\put(90,108){\line(1,0){30}}
\put(120,109){\line(1,0){30}}
\put(180,109){\vector(-1,0){30}}
\put(180,109){\vector(1,0){30}}
\put(210,109){\line(1,0){30}}
\put(120,107){\line(1,0){30}}
\put(180,107){\vector(-1,0){30}}
\put(180,107){\vector(1,0){30}}
\put(210,107){\line(1,0){30}}
\put(240,108){\line(1,0){30}}
\put(300,108){\vector(-1,0){30}}
\put(120,84){\line(0,1){24}}
\put(240,84){\line(0,1){24}}
\put(116,81){$\times$}
\put(176,105){$\times$}
\put(236,81){$\times$}
\put(75,117){$\ell^+_{(16)}$}
\put(140,117){$\ell^{- \prime}_{ (16')}$}
\put(197,117){$\overline{\ell^-}'_{(\overline{16}')}$}
\put(262,117){$\ell^-_{(16)}$}
\put(102,66){$\underbrace{\langle H_{(10_H)} \rangle}$}
\put(170,88){$M^{\dag}$}
\put(230,66){$\underbrace{\langle \Omega \rangle}$}
\put(113,40){$m_H$}
\put(233,40){$M_{\Omega}$}
\put(168,0){{\bf Fig. 1(a)}}
\end{picture}

\begin{picture}(360,216)
\thicklines
\put(60,108){\vector(1,0){30}}
\put(90,108){\line(1,0){30}}
\put(120,109){\line(1,0){30}}
\put(180,109){\vector(-1,0){30}}
\put(180,109){\vector(1,0){30}}
\put(210,109){\line(1,0){30}}
\put(120,107){\line(1,0){30}}
\put(180,107){\vector(-1,0){30}}
\put(180,107){\vector(1,0){30}}
\put(210,107){\line(1,0){30}}
\put(240,108){\line(1,0){30}}
\put(300,108){\vector(-1,0){30}}
\put(120,84){\line(0,1){24}}
\put(240,84){\line(0,1){24}}
\put(116,81){$\times$}
\put(176,105){$\times$}
\put(236,81){$\times$}
\put(75,117){$\ell^+_{(16)}$}
\put(140,117){$\overline{\ell^+}'_{ (\overline{16}')}$}
\put(197,117){$\ell^{+\prime}_{ (16')}$}
\put(262,117){$\ell^-_{ (16)}$}
\put(110,66){$\underbrace{\langle \Omega \rangle}$}
\put(170,88){$M^*$}
\put(222,66){$\underbrace{\langle H_{(10_H)} \rangle}$}
\put(113,40){$M_{\Omega}^T$}
\put(233,40){$m_H^T$}
\put(168,0){{\bf Fig. 1(b)}}
\end{picture}

\newpage 

\begin{picture}(360,216)
\thicklines
\put(60,108){\vector(1,0){30}}
\put(90,108){\line(1,0){30}}
\put(120,109){\line(1,0){30}}
\put(180,109){\vector(-1,0){30}}
\put(180,109){\vector(1,0){30}}
\put(210,109){\line(1,0){30}}
\put(120,107){\line(1,0){30}}
\put(180,107){\vector(-1,0){30}}
\put(180,107){\vector(1,0){30}}
\put(210,107){\line(1,0){30}}
\put(240,108){\line(1,0){30}}
\put(300,108){\vector(-1,0){30}}
\put(120,84){\line(0,1){24}}
\put(240,84){\line(0,1){24}}
\put(116,81){$\times$}
\put(176,105){$\times$}
\put(236,81){$\times$}
\put(75,117){$N^c_{(16)}$}
\put(140,117){$\nu'_{ (16')}$}
\put(197,117){$\overline{\nu}'_{(\overline{16}')}$}
\put(262,117){$\nu_{(16)}$}
\put(102,66){$\underbrace{\langle H_{(10_H)} \rangle}$}
\put(170,88){$M^{\dag}$}
\put(232,66){$\underbrace{\langle \Omega \rangle}$}
\put(113,40){$m_H$}
\put(233,40){$M_{\Omega}$}
\put(168,0){{\bf Fig. 2(a)}}
\end{picture}

\begin{picture}(360,216)
\thicklines
\put(60,108){\vector(1,0){30}}
\put(90,108){\line(1,0){30}}
\put(120,109){\line(1,0){30}}
\put(180,109){\vector(-1,0){30}}
\put(180,109){\vector(1,0){30}}
\put(210,109){\line(1,0){30}}
\put(120,107){\line(1,0){30}}
\put(180,107){\vector(-1,0){30}}
\put(180,107){\vector(1,0){30}}
\put(210,107){\line(1,0){30}}
\put(240,108){\line(1,0){30}}
\put(300,108){\vector(-1,0){30}}
\put(120,84){\line(0,1){24}}
\put(240,84){\line(0,1){24}}
\put(116,81){$\times$}
\put(176,105){$\times$}
\put(236,81){$\times$}
\put(75,117){$N^c_{(16)}$}
\put(140,117){$\overline{N^c}'_{(\overline{16}')}$}
\put(197,117){$N^{c \prime}_{(16')}$}
\put(262,117){$\nu_{(16)}$}
\put(112,66){$\underbrace{\langle \Omega \rangle}$}
\put(170,88){$M^*$}
\put(222,66){$\underbrace{\langle H_{(10_H)} \rangle}$}
\put(113,40){$M_{\Omega}^T$}
\put(233,40){$m_H^T$}
\put(168,0){{\bf Fig. 2(b)}}
\end{picture}

\newpage 

\begin{picture}(360,216)
\thicklines
\put(0,108){\vector(1,0){25}}
\put(25,108){\line(1,0){20}}
\put(45,109){\line(1,0){20}}
\put(90,109){\vector(-1,0){25}}
\put(90,109){\vector(1,0){25}}
\put(115,109){\line(1,0){20}}
\put(45,107){\line(1,0){20}}
\put(90,107){\vector(-1,0){25}}
\put(90,107){\vector(1,0){25}}
\put(115,107){\line(1,0){20}}
\put(135,108){\line(1,0){20}}
\put(180,108){\vector(1,0){25}}
\put(180,108){\vector(-1,0){25}}
\put(205,108){\line(1,0){20}}
\put(225,109){\line(1,0){20}}
\put(270,109){\vector(-1,0){25}}
\put(270,109){\vector(1,0){25}}
\put(295,109){\line(1,0){20}}
\put(225,107){\line(1,0){20}}
\put(270,107){\vector(-1,0){25}}
\put(270,107){\vector(1,0){25}}
\put(295,107){\line(1,0){20}}
\put(315,108){\line(1,0){20}}
\put(360,108){\vector(-1,0){25}}
\put(45,84){\line(0,1){24}}
\put(135,84){\line(0,1){24}}
\put(180,84){\line(0,1){24}}
\put(225,84){\line(0,1){24}}
\put(315,84){\line(0,1){24}}
\put(41,81){$\times$}
\put(131,81){$\times$}
\put(176,81){$\times$}
\put(221,81){$\times$}
\put(311,81){$\times$}
\put(86,105){$\times$}
\put(266,105){$\times$}
\put(12,117){$\nu_{(16)}$}
\put(60,117){$N^{c \prime}_{(16')}$}
\put(100,117){$\overline{N^c}'_{(\overline{16}')}$}
\put(150,117){$N^c_{(16)}$}
\put(195,117){$N^c_{(16)}$}
\put(235,117){$\overline{N^c}'_{(\overline{16}')}$}
\put(285,117){$N^{c \prime}_{(16')}$}
\put(330,117){$\nu_{(16)}$}
\put(29,66){$\underbrace{\langle H_{(10_H)} \rangle \hskip0.9in 
\langle \Omega \rangle}$}
\put(86,88){$M^{\dag}$}
\put(163,66){$\underbrace{\langle \overline{126}_H \rangle}$}
\put(220,66){$\underbrace{\langle \Omega \rangle \hskip0.9in
\langle H_{(10_H)} \rangle}$}
\put(266,88){$M^*$}
\put(76,44){$\subset M_N$}
\put(172,44){$M_R$}
\put(266,44){$\subset M_N^T$}
\put(168,0){{\bf Fig. 3}}
\end{picture}

\vspace{1cm}

\begin{picture}(360,216)
\thicklines
\put(0,108){\vector(1,0){30}}
\put(30,108){\line(1,0){30}}
\put(60,109){\line(1,0){30}}
\put(120,109){\vector(-1,0){30}}
\put(120,109){\vector(1,0){30}}
\put(150,109){\line(1,0){30}}
\put(180,109){\line(1,0){30}}
\put(240,109){\vector(1,0){30}}
\put(240,109){\vector(-1,0){30}}
\put(270,109){\line(1,0){30}}
\put(60,107){\line(1,0){30}}
\put(120,107){\vector(-1,0){30}}
\put(120,107){\vector(1,0){30}}
\put(150,107){\line(1,0){30}}
\put(180,107){\line(1,0){30}}
\put(240,107){\vector(1,0){30}}
\put(240,107){\vector(-1,0){30}}
\put(270,107){\line(1,0){30}}
\put(300,108){\line(1,0){30}}
\put(360,108){\vector(-1,0){30}}
\put(60,84){\line(0,1){24}}
\put(180,84){\line(0,1){24}}
\put(300,84){\line(0,1){24}}
\put(56,81){$\times$}
\put(116,105){$\times$}
\put(176,81){$\times$}
\put(236,105){$\times$}
\put(296,81){$\times$}
\put(20,117){$\nu_{(16)}$}
\put(80,117){$N^{c \prime}_{(16')}$}
\put(140,117){$\overline{N^c}'_{(\overline{16}')}$}
\put(200,117){$\overline{N^c}'_{(\overline{16}')}$}
\put(260,117){$N^{c \prime}_{(16')}$}
\put(320,117){$\nu_{(16)}$}
\put(39,66){$\underbrace{\langle H_{(10_H)} \rangle}$}
\put(110,88){$M^{\dag}$}
\put(164,66){$\underbrace{\langle 126_H \rangle}$}
\put(230,88){$M^*$}
\put(276,66){$\underbrace{\langle H_{(10_H)} \rangle}$}
\put(54,44){$m_H$}
\put(174,44){$\overline{M}'_R$}
\put(294,44){$m^T_H$}
\put(168,0){{\bf Fig. 4}}
\end{picture}

\newpage 

\begin{picture}(360,216)
\thicklines
\put(60,108){\vector(1,0){30}}
\put(90,108){\line(1,0){30}}
\put(120,109){\line(1,0){30}}
\put(180,109){\vector(-1,0){30}}
\put(180,109){\vector(1,0){30}}
\put(210,109){\line(1,0){30}}
\put(120,107){\line(1,0){30}}
\put(180,107){\vector(-1,0){30}}
\put(180,107){\vector(1,0){30}}
\put(210,107){\line(1,0){30}}
\put(240,108){\line(1,0){30}}
\put(300,108){\vector(-1,0){30}}
\put(120,84){\line(0,1){24}}
\put(180,84){\line(0,1){24}}
\put(240,84){\line(0,1){24}}
\put(116,81){$\times$}
\put(176,81){$\times$}
\put(236,81){$\times$}
\put(80,117){$\nu_{(16)}$}
\put(140,117){$N^{c \prime}_{(16')}$}
\put(200,117){$N^{c \prime}_{(16')}$}
\put(260,117){$\nu_{(16)}$}
\put(102,64){$\langle H_{(10_H)} \rangle$}
\put(162,64){$\langle \overline{126}_H \rangle$}
\put(222,64){$\langle H_{(10_H)} \rangle$}
\put(168,0){{\bf Fig. 5}}
\end{picture}


\begin{thebibliography}{999}
\bibitem{seesaw} M. Gell-Mann, P. Ramond, and R. Slansky, in 
{\it Supergravity}, Proceedings of the Workshop. Stony Brook, New York, 1979, 
ed. P. van Nieuwenhuizen and D.Z. Freedman (North-Holland, Amsterdam, 1979),
p. 315; T. Yanagida, in {\it Proc. Workshop on Unified Theory and the
Baryon Number of the Universe}, Tsukuba, Japan, 1979, ed. O. Sawada
and A. Sugramoto (KEK Report No. 79-18, Tsukuba, 1979); R.N. Mohapatra
and G. Senjanovic, {\it Phys. Rev. Lett.} {\bf 44}, 912 (1980); S.L. Glashow
in {\it Quarks and leptons}, Cargese (July 9-29, 1979), ed. M. Levy et al.
(Plenum, New York, 1980), p. 707.
\bibitem{bd} S.M. Barr and I. Dorsner, {\it Nucl. Phys.} {\bf B585},
79 (2000).
\bibitem{atm} Y. Ashie et al., (Super-Kamiokande Collaboration), 
hep-ex/0404034; E. Kearns (Super-Kamiokande Collaboration), talk given
at Neutrino2004.
\bibitem{solar} T. Araki et al., (KamLAND Collaboration) hep-ex/0406035;
G. Gratta (KamLAND Collaboration), talk given at Neutrino2004;
J.N. Bahcall and C. Pena-Garay, JHEP 0311:004 (2003).
\bibitem{fn} C.D. Froggatt and H.B. Nielson, {\it Nucl. Phys.} {\bf B147},
277 (1979).
\bibitem{models} S.M. Barr, {\it Phys. Rev.} {\bf D21}, 1424 (1980);
R. Barbieri and D.V. Nanopoulos, {\it Phys. Lett.} {\bf B95}, 43 (1980);
S.M. Barr, {\it Phys. Rev.} {\bf D24}, 1895 (1981); {\it Phys. Rev.}
{\bf D42}, 3150 (1990); G. Anderson, S. Dimopoulos, L.J. Hall, S. Raby, and 
G.D. Starkman, {\it Phys. Rev.} {\bf D49}, 3660 (1994); C.H. Albright 
and S.M. Barr, {\it Phys. Rev.}; C.H. Albright, K.S. Babu, and S.M. Barr, 
{\it Phys. Rev. Lett.} {\bf 81}, 1167 (1998); K.S. Babu, J. Pati, and F. Wilczek,
{\it Nucl. Phys.} {\bf B566}, 33 (2000).
\bibitem{gen} S.M. Barr and B. Kyae, {\it Phys. Rev.} {\bf D70},
075005 (2004).
\bibitem{ns} Y. Nir and Y. Shadmi, JHEP 0411:055 (2004).
\end{thebibliography}
\end{document}